 \documentclass{elsart}

 \usepackage{graphicx}
\usepackage{amssymb}

\begin{document}

\begin{frontmatter}



\title{SUMMARY OF 
EXPERIMENTAL STUDIES, AT CERN, ON A POSITRON SOURCE USING CRYSTAL EFFECTS
\thanksref{titleref}
}
\thanks[titleref]{Research under INTAS contract 97-562}


\author[IPN-Lyon]{X.Artru},
\author[BINP]{V.Baier},
\author[BINP]{K.Beloborodov},
\author[NPI-TPU]{A.Bogdanov},
\author[BINP]{A.Bukin},
\author[BINP]{S.Burdin\thanksref{burdin}},
\thanks[burdin]{presently at FNAL, Batavia, IL, USA}
\author[LAL]{R.Chehab\corauthref{cor}},
\ead{chehab@lal.in2p3.fr}
\corauth[cor]{Corresponding author.}
\author[IPN-Lyon]{M.Chevallier},
\author[LAL]{R.Cizeron},
\author[IPN-Lyon]{D.Dauvergne},
\author[BINP]{T.Dimova},
\author[BINP]{V.Druzhinin},
\author[BINP]{M.Dubrovin\thanksref{dubrovin}}, 
\thanks[dubrovin]{presently at Wayne State University - Cornell, Ithaca NY, USA}
\author[CERN]{L.Gatignon},
\author[BINP]{V.Golubev},
\author[LMD]{A.Jejcic},
\author[Max-Planck]{P.Keppler},
\author[IPN-Lyon]{R.Kirsch},
\author[KIPT]{V.Kulibaba},
\author[IPN-Lyon]{Ph.Lautesse},
\author[Max-Planck]{J.Major},
\author[IPN-Lyon]{J-C.Poizat},
\author[NPI-TPU]{A.Potylitsin},
\author[IPN-Lyon]{J.Remillieux},
\author[BINP]{S.Serednyakov},
\author[LAL,BINP]{V.Shary},
\author[BINP]{V.Strakhovenko},
\author[LAL]{C.Sylvia},

\address[LAL]{LAL, IN2P3-CNRS and Universit\'e de Paris-Sud, 91898 Orsay Cedex, France}
\address[BINP]{BINP, Av.~Lavrentyeva, 11, 630090 Novosibirsk, Russia}
\address[IPN-Lyon]{IPN-Lyon, IN2P3-CNRS and Univ. Claude Bernard, 69622
Villeurbanne, France}
\address[CERN]{CERN, CH-1211 Gen\`eve 23, Switzerland}
\address[Max-Planck]{Max-Planck Institute, Heisenberg Str.~70569 Stuttgart, Germany}
\address[NPI-TPU]{NPI-TPU, Prospekt Lenina, 634050 Tomsk, Russia}
\address[KIPT]{KIPT, Academisheskaya Str. 310108 Kharkov, Ukraine}
\address[LMD]{LMD, Universit\'e Paris, Jussieu 75005-Paris, France }

\begin{abstract}
A new kind of positron sources for future linear colliders, 
where the converter is 
an aligned tungsten crystal, oriented on the $\left<111\right>$ axis,
has been studied at CERN in the WA103 experiment with tertiary electron beams from the SPS.
In such sources the photons resulting from channeling 
radiation and coherent bremsstrahlung create the $e^+ e^-$ pairs.

Electron beams, of 6 and 10 GeV, 
were impinging on different kinds of targets: 
a 4 mm thick crystal, a 8 mm thick crystal 
and a compound target made of 4 mm crystal followed by 4 mm amorphous disk. 
An amorphous tungsten target 20 mm thick was
also used for the sake of comparison with the 8 mm crystal and to check the
ability of the detection system to provide the correct track reconstruction.
The charged particles coming out from the target were detected in
a drift chamber immersed partially in a magnetic field. The reconstruction
of the particle trajectories provided the energy and angular spectrum of the
positrons in a rather wide energy range (up to 150 MeV) and angular domain
(up to 30 degrees). The experimental approach presented in this article 
provides a full
description of this kind of source. A presentation of the measured
positron distribution in momentum space (longitudinal versus transverse) 
is given to allow an easy
determination of the available yield for a given momentum acceptance.
Results on photons, measured downstream of the positron detector, are also
presented.
  A significant
enhancement of photon and positron production is clearly observed. 
This enhancement, for a 10 GeV
incident beam, is of 4 for the 4 mm thick crystal and larger than 2 for the 8 mm
thick crystal.
Another important result concerns the
validation of the simulations for the crystals, for which a quite good
agreement was met between the simulations and the experiment, for positrons as 
well as for photons.
  These results are presented after a short presentation of the experimental
setup and of the track reconstruction procedure.
\end{abstract}

\end{frontmatter}

\section{Introduction}
Linear electron-positron colliders will be the privileged tool for studying physics in
the TeV energy region. To overcome the decrease of the $e^+e^-$ annihilation cross
section, very large currents both in $e^-$ and in $e^+$ will be needed. The conventional
positron source consists in an amorphous target of heavy metal, hit by a primary
electron beam. Photons, which are produced by bremsstrahlung, are converted  in the same
target into $e^+e^-$ pairs resulting in an electromagnetic shower. The thickness of the target is chosen to maximize 
the number of the positrons in the energy
and angular domain of acceptance of the downstream matching system installed before the
damping ring. Such positrons represent only a small fraction of all charged particles created 
in the target. This is due to their wide energy and angular spread, which comes from the
photon emission and pair production processes and, above all, from  multiple scattering of charged particles. 
Therefore a very intense primary electron beam is needed to achieve the desired positron beam current.

In the high energy region, the basic processes involved in the shower
development are considerably enhanced in the oriented crystals, as
compared with corresponding amorphous media. The
most pronounced effects take place at the orientation where the electrons
are incident along the major axes of the crystal. At such orientation, the
incident electrons are exposed to the highest mean electric field provided
by the atomic rows \cite{BKS2}. The enhancement
depends on the particle energy and the crystal orientation (see \cite{BKS2a}
for further details concerning electromagnetic processes in crystals). 
Therefore, the replacement of the conventional positron
target by a crystal is expected to improve the efficiency of the positron source.

When the energy increases, the coherent (crystal) effects become noticeable 
first in radiation, while their contribution 
to the pair production is still negligible. For example, the radiation 
intensity  caused by the electric field of 
the $ <111> $-axis of a tungsten crystal exceeds that of the conventional 
bremsstrahlung starting with electron energies 
$\varepsilon \sim $ 1 GeV \cite{BKS2}.
% In the high energy region, the basic processes involved in the shower
% development are considerably enhanced in the oriented crystals, as
% compared with corresponding amorphous media. The enhancement
% depends on the particle energy and the crystal orientation (see \cite{BKS2a} and \cite{BKS2}
% for further details concerning electromagnetic processes in crystals). The
% most pronounced effects take place at the orientation where the electrons
% are incident along the major axes of the crystal. At such orientation, the
% incident electrons are exposed to the highest mean electric field provided
% by the atomic rows. Therefore, the replacement of the conventional positron
% target by a crystal is expected to improve the efficiency of the positron source.
% When the energy increases, the coherent (crystal) effects become noticeable first in radiation, while their contribution 
% to the pair production is still negligible. For example, the radiation intensity  caused by the electric field of 
% the $ <111> $-axis of a tungsten crystal exceeds that of the conventional bremsstrahlung starting with electron energies 
% $\varepsilon \sim $ 1 GeV. 
For the same crystal, the coherent contribution to the pair production starts to exceed that 
 of the standard (Bethe-Heitler) mechanism at photon energies $\omega\sim $ 22 GeV. For lighter crystals, the 
 corresponding value of $\omega$ turns out to be several times larger than for tungsten. An incident electron energy 
 of several GeV is foreseen for an efficient positron source. Then the pair production process proceeds in a crystal as in 
 the corresponding amorphous medium. The enhancement of radiation from 
 initial electrons is thereby the main crystal effect in the energy region of interest. 
According to
theoretical estimates (see, e.g.~\cite{BKS2}), the effective radiation length $L_{ef}$ in a 
tungsten crystal becomes smaller than the conventional radiation length by the factor of 4.2 at 
$\varepsilon = $ 4 GeV and by the factor of 5.7 at $\varepsilon = $ 8 GeV. Alongside with the high power, the radiation at 
axial alignment is characterized by the softness of its spectrum. This leads to further increase in the number of emitted photons 
and in the number of produced positrons as compared with the amorphous target. 

The description of the development of the electron-photon showers in an axially aligned crystal was elaborated in \cite{Artr}, 
\cite{BKS4}, \cite{BS1}. According to these papers (see also Fig.~1 in \cite{ACCS}), the electron energy is converted in a crystal 
target
into photons over a thickness of several  $L_{ef}$. Then  at the depth  $L_0 \approx (3 \div 4) L_{ef}$ most of the particles,
including the initial electrons, are sufficiently soft to reduce the coherent contribution to the radiation to a level 
below the
incoherent one. Therefore, the further development of the shower proceeds more or less in the same way for the 
crystal or amorphous
type of the remaining part of a target. So, we come to the idea of a compound target, which consists of a 
crystal part of thickness about $L_0$ followed by an amorphous one
\cite{Chehab}.
We emphasize that the crystal part of a target serves as a radiator, and
secondary charged particles are still not so numerous at this stage of the shower development. Therefore, only a small portion
of the total energy loss is deposited in this part of the target, which considerably reduces the danger of its overheating. 

In the construction of an intense positron source, the thermal issues are rather important. A substantial part (for example, 
more than 30  \%  in the JLC project~\cite{JLC}) of the primary beam power is dissipated and transformed into heat inside the target, 
mainly due to the ionization energy losses by the charged particles of the shower. As already observed in the SLAC Linear Collider (SLC) target, this causes 
serious thermal problems, which makes the actual limitation to the beam current. For crystals, where the ionization energy losses are
practically not modified as compared with amorphous media, the thermal issues were considered in  \cite{BKS4}, \cite{BS1} and, in
detail, in  \cite{ACCS}. It was found that, in the energy range under consideration, the positron yield at the optimal target 
thickness may be larger in a crystal case only by several percent. 
However, the amount of the energy deposition in a crystal 
turns out to be considerably lower (by a factor 2 for the JLC project, for instance) 
than in an amorphous target providing the same positron yield, while the
Peak Energy Deposition Density (PEDD) is approximately of the same magnitude in both cases.

Recently the positron production in axially aligned  single crystals was studied in two series of experiments performed at 
CERN  \cite{C01} and KEK  \cite{O01}. In both cases, the incident electron beam of different energies, up to  10~GeV,  was 
aligned with the $<111>$-axis of a tungsten crystal, which sometimes served as a crystal part of the compound target 
containing an additional amorphous tungsten part. Basing on the approach developed in \cite{BKS4} and \cite{BS1}, theoretical 
estimations were obtained in \cite{BS2}, which display a rather good agreement with the experimental results of \cite{O01},
where only the positrons outgoing in the forward direction (i.e., with a vanishing transverse momentum) were registered for a 
discrete set of their longitudinal momentum values. However, to obtain a full description of the crystal-assisted positron 
source, the positron distribution in the whole momentum space, longitudinal and transverse, should be measured. This is done in
the present experimental studies, where the energy and angular distributions of the positrons are obtained in a rather wide 
energy range (up to 150 MeV) and angular domain (up to 30 degrees).

\section{The experimental setup}

The experiment was installed on the X5 transfer line of the SPS West Hall at 
CERN. 
It used tertiary electron beams having energies between 5 and 40 GeV. 
After passing through trigger counters and profile monitors (delay 
chambers) the electrons impinge on the target. Photons and $e^+e^-$ pairs are 
generated in this target. They go mainly in the forward direction and 
travel across the magnetic spectrometer made of a
magnet (MBPS), with vertical magnetic field,
inside which the drift chamber is placed.
The drift chamber is completed 
%(flanked??) 
by two positron counters, 
one on the wall away from the beam, one on the back wall (Fig.~\ref{set-up}). 
The charged particles coming out from the Drift Chamber 
close to the forward direction 
are swept by a second magnet (MBPL) while the forward photons reach the photon detector 
made of a preshower and a calorimeter. 
 
\subsection{The beam} 
  The SPS beam is made of 3.2 (resp. 5.2) second duration pulses 
with a 14.4 (resp. 16.8) second period for the 2000 (resp. 2001) summer runs. 
Each pulse contains $\sim 10^4$ electrons 
representing 99~\% of the particles. This intensity value was currently  
reached at 6 and 10 GeV. 

The channeling condition requires that the incident electron angle on 
the target be smaller than the Lindhard critical angle \cite{LA}; this angle is of 
0.45 mrad at 10 GeV and for the $\left<111\right>$ orientation of the tungsten crystal. 
Taking into account the persistence of crystal effects at 
angles larger than the critical angle, the acceptance angle for the trigger 
was chosen equal to 0.75 mrad. This selects about 1\% of the incident electrons.
A trigger system made of scintillation counters, 
with a restricted angular aperture of 0.75 mrad, 
was installed upstream of the target. 
The trigger selection was 
improved off line by the informations provided by a proportional delay chamber.  
This delay chamber with horizontal and vertical grids, put at  
short distance from the target, provided informations on the lateral  
distributions of the incident electron beam. These informations served 
for beam control during  the run and, as said above,  
were also used in the selection criteria.  
On Fig.~\ref{delay} 
we represent the lateral distributions of the electron beam.

\subsection{The targets} 

 The crystal-assisted positron source can be an all-crystal target. However, 
coherent effects are mostly important at the beginning of the shower. In the 
following steps of the shower the particles  have lower energies, larger 
angles and the crystalline effects are not important. Therefore, a 
compound target, made of a crystal as a primary radiator generating a large 
amount of photons and an amorphous disk where the photons are converted into $e^+e^-$ pairs, has nearly the same advantages. In the experiment, four kinds of 
targets have been installed on a 0.001 degree precision goniometer: 
\begin{itemize}
\item  a 4 mm thick crystal, 
\item  a 8 mm thick crystal, 
\item  a compound target made of 4 mm crystal followed by a 4 mm thick 
          amorphous disk, 
\item  an amorphous disk 20 mm thick; this target was used in order to 
            check the reconstruction efficiency with a number of 
            positrons comparable to the one produced in a 8 mm thick aligned crystal. 
\end{itemize}

  Mosaic spreads of the crystals (root mean square disorientation angle of the crystalline
   domains) is measured by gamma-diffractometry at the 
  Max-Planck Institute in Stuttgart- are slightly less than 0.5 mrad. The targets 
  alignment was done using the data from the positron counters and from the 
  photon preshower. The latter gives a signal proportional to the number of photons, 
  which is maximum on axis position, as can be seen on the rocking curves 
(Fig.~\ref{scan}).  
The signal in the preshower is formed mainly by the two processes: $\gamma$ conversion
with the yield of two minimal ionizing particles (MIP) per photon and Compton scattering 
with the yield of one MIP per photon. The relative contributions of the processes depend on the emitted 
$\gamma$ spectrum.

\subsection{ The positron detector} 
  
\begin{figure*}
\centerline{\includegraphics[width=\textwidth]{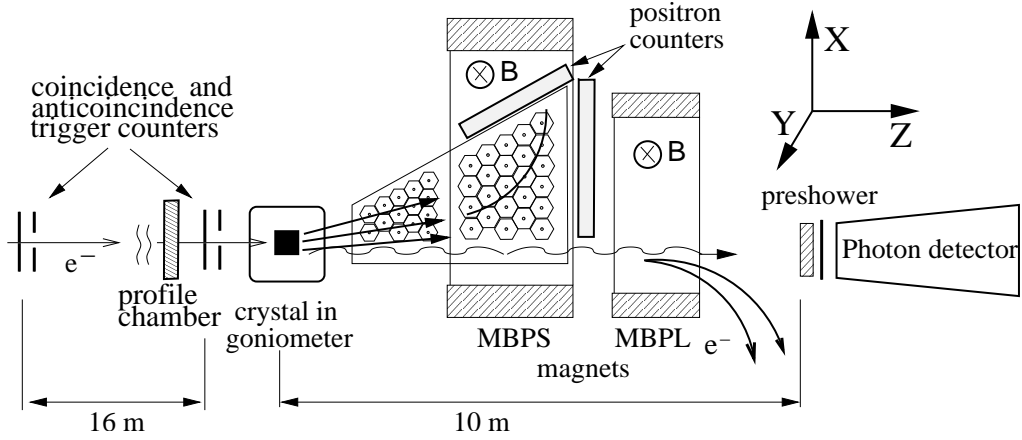}}
\caption{The set-up scheme.
Drift chamber in two parts. DC1 is outside the magnetic field. 
DC2 is in the magnetic field of MBPS magnet. MBPL is the sweeping magnet. 
\label{set-up}}
\end{figure*}

 The positron detector consists in the drift chamber (DC) with hexagonal cells 
and positron scintillation counters. The DC is a quasi-plane detector with its
largest dimensions in the horizontal plane and the smallest one, in the vertical
plane. It presents two parts: 
\begin{itemize}
\item  the first part (DC1), with a cell radius  of 0.9 cm, is located mainly
         outside the magnetic field of the spectrometer. It allows the
	 measurement of the horizontal exit angle, $\alpha$ of the positrons. 
\item  the second one (DC2), with a cell radius of 1.6 cm, is immersed
         in the magnetic field. It allows the measurement of the horizontal 
	 positron 
         (electron) momentum. Two values of the magnetic field are used: 
	 1 and 4 kilogauss (KG), in order to investigate the whole energy region of 
         interest, from 5 to 150 MeV.
\end{itemize}	 
In the drift chamber the wires are parallel to the magnetic field (vertical).
 The available space in the bending magnet (MBPS) restricts the vertical size of the 
chamber and therefore, the  length of the wires (6 cm).	  
  With such short length, the border effect is significant. 
 The drift chamber has metallic walls therefore these border effects lead to 
an increase of the gas amplification. In order to use the central part  
of the wires, two positron counters with a vertical size of 3 cm were used. 
That sets the vertical acceptance to $\pm 1.5$ degree
for a distance target -- counter of about 1 meter.
The first counter is 
placed on the lateral side of the chamber and the other one on the back side. 
(see Fig.~\ref{set-up}). Low energy positrons are more likely bent toward the side 
positron counter. Strictly speaking, for a many-track event, these counters 
told only that at least one track was hitting the central part of the wires 
but did not tell which one. The signal provided by the side counter gives a 
rather good indication on the low-energy positron yield. 
 The signal wires in the drift chamber are made of gold-plated tungsten and 
the field wires of titanium. The drift chamber has 21 layers. The maximum  
horizontal angle being accepted is 30 degrees. The limited vertical size 
sets the overall acceptance to 6 \% of $2\pi$ solid angle. 
In order to reduce the multiple 
scattering for low-energy positrons the drift chamber is filled with an 
helium-based mixture {He (90 \%) + CH$_4$ (10 \%)}. The achieved coordinate resolution 
is about 500 micrometers. This leads to an angular resolution of 0.25 
degree and a typical energy resolution of 0.6 MeV for both values of the 
magnetic field. 
 Standard electronics associated to drift chambers is used. The data 
acquisition software is developed using KMAX system under Windows NT 
environment. The collected data is saved on a hard disk. The total amount 
of the raw data is approximately 3 GB. 
 
\subsection{ The photon detector} 
    
  The photon multiplicity is rather high: about 200 photons/event for a 8 mm 
thick crystal target oriented along its $<111>$ axis and 10 GeV initial 
electron energy. 
We measure only the average photon multiplicity
and the total radiated energy, contained in a forward cone with maximum half
angle of 4 mrad. For this goal the
photon detector is made of:
\begin{itemize}
\item a preshower made of a copper disk ($0.2 X_0$ thick), followed by a 
       scintillator. Two sizes of copper disks (3 and 6 cm in diameter, 
       corresponding to a semi-angle aperture of 1.5 and 3 mrad respectively) 
       have been used successively to look at the number of photons and at  
       their angular distribution. The preshower gives information on the 
       average photon 
       multiplicity. As the number of photons varies more sharply than  
       the radiated energy with respect to the crystal orientation, 
       the preshower signal is used for the crystal alignment. 
\item  a "spaghetti" calorimeter with thin scintillation fibers, giving the 
       amount of radiated energy (see~\cite{Bellini}). The preshower is in close 
       contact with the "spaghetti" calorimeter.     
\end{itemize} 
  
\begin{figure}
\includegraphics[width=.48\textwidth]{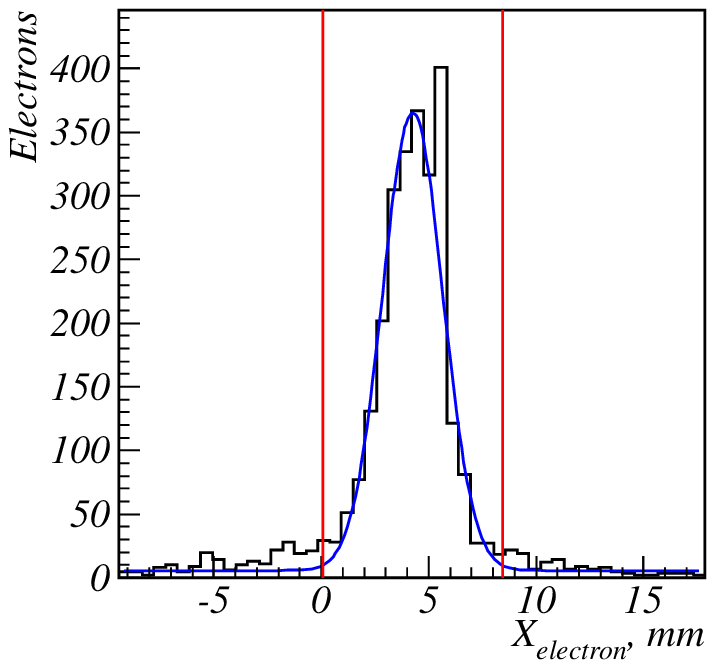}
\hfill
\includegraphics[width=.48\textwidth]{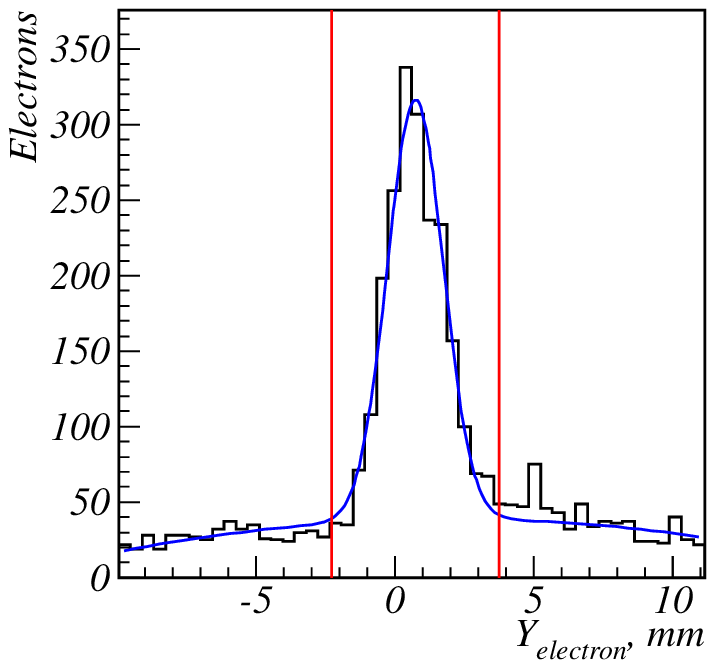}
\caption{The x and y coordinate of the incident electron measured by the profile 
monitor. Lines show the cuts used in the analysis.
\label{delay}}
\end{figure}

\begin{figure}
\includegraphics[width=.48\textwidth]{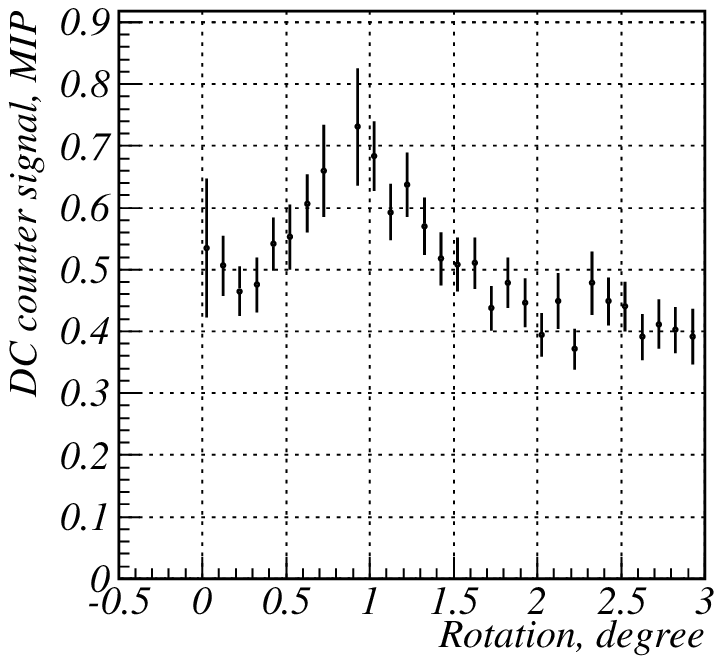}
\hfill
\includegraphics[width=.48\textwidth]{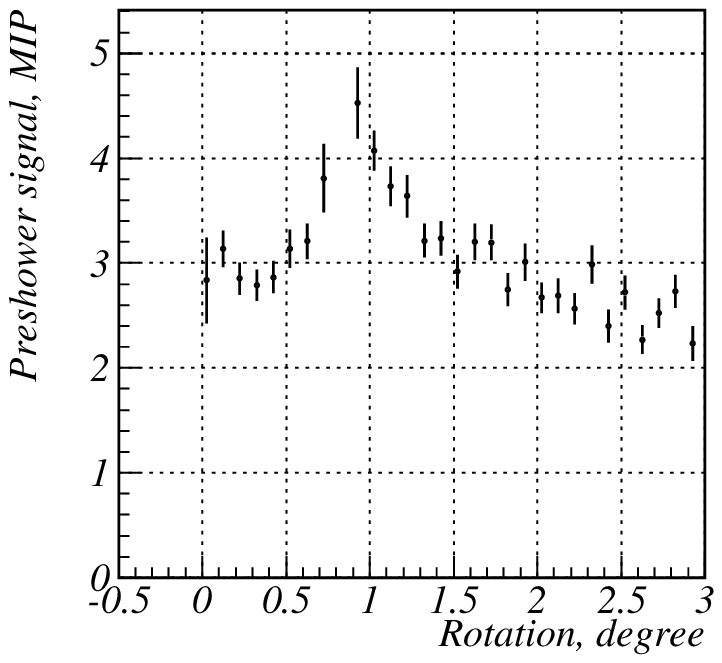}
\caption{Average amplitudes in the counters as a function of the rotation angle 
(rocking curves). Left figure is the signal in the positron counter and the right one is the 
signal in the preshower. 
\label{scan}}
\end{figure}

\section{Analysis}

\subsection{The reconstruction procedure}
The track reconstruction was done with the histogram method, modified 
to assign 3 parameters to each track. 
To avoid difficulties
coming from the high rate of occupancy and complications from the
non-uniform magnetic field in the first part of the drift chamber, the
reconstructed trajectory was based on the informations from the second part
of the chamber. The extrapolation back to the target has been operated using
 the measured values of the field.
The trajectory of an electron or positron, projected on the horizontal plane, 
was parametrised as a circle with the following parameters:
$r$ is the radius of the circle; $(x_c, z_c)$ are the coordinates of the circle center.
To determine the track parameters, a minimum of 3 hits must be detected.
Each hit may be represented by a small circle centered on the hit wire and 
the radius of which is proportional to the drift time.
If {\it n} hits were detected in the drift chamber, the number of all possible tracks
based on 3 hits each is $N=\frac{8 n!}{3! (n-3)!}$, according to
combinatorial analysis.
The factor 8 comes from the two possible contact points between the track and 
the small hit circle, this choice being repeated 3 times. 
For example for the 20 mm target and 10 GeV incident electron the typical
number of hits is near 50 and the typical number of reconstructed tracks is 4 -- 5.
All these tracks parameters are collected in three 2D-histograms: 
($x_c, z_c$), ($x_c, r$) and ($z_c, r$). 
All the hit triplets produced by the same real track will give approximately 
the same values of parameters and hence  will produce a peak on the histograms.
The histogram peak position serve as a parameter seed for the fitting procedure. 
The candidate with maximal number of hits is selected and all the hits belonging 
to this candidate are removed from the further consideration.
After that, the next track is chosen with the same procedure and this procedure is
repeated as long as at least one candidate with 4 hits exists.
For each track found the optimal set of parameters is calculated:
the charge $Q$, 
the horizontal momentum $p_h=\sqrt{p_x^2+p_z^2}$ 
%is a momentum projection on the horizontal plane, 
%perpendicular to the magnetic field,
and the horizontal angle $\alpha = \arctan{(p_x/p_z)}$ 
with respect to the beam axis.
%!!OUBLIER "$x_t$ is a coordinate..."
% of the trajectory projection on the target plane.

\subsection{Reconstruction efficiency}

\begin{figure}
\includegraphics[width=.5\textwidth]{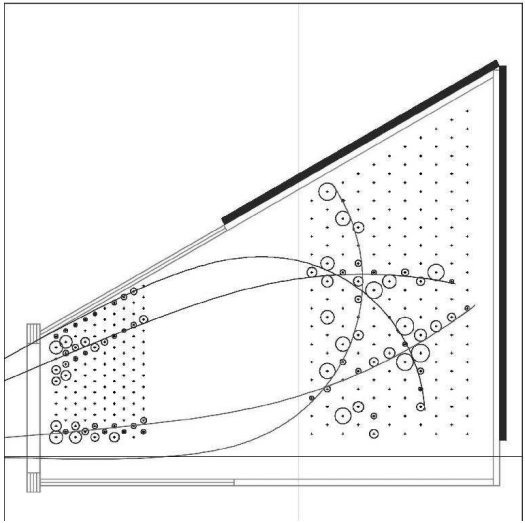}
\caption{A typical reconstructed event for the 20 mm target, the electron
energy is 10 GeV.
The crosses are the drift chamber wires. Circles around the crosses represent the
measured distance between tracks and wires. Lines are the reconstructed electron
 and positrons trajectories.
\label{event}}
\end{figure}

We define a reconstruction efficiency $\eta(p,\theta)$ as:
\begin{equation} 
\eta(p, \theta) = {N_r^{sim} \over N_{in}^{sim}} \,,
\end{equation}
where $N_r^{sim}$ is the number of reconstructed tracks,
$N_{in}^{sim}$, the number of positrons passing through the chamber, i.e.,
crossing a minimum number of cells and where $p_h$ is taken for $p$ and $\alpha$ for the $\theta$
due to the flatness of the drift chamber.
This efficiency is determined by the simulations. It has been calculated
with GEANT code \cite{Geant}. For the case with high occupancy (20mm
amorphous target and 10 GeV incident electron beam) it reaches 80\% at
large angles. 
However for the tracks with angles $\alpha < 7$ degrees (usually high
positron momenta), this efficiency is reduced to 50 \% due to the high
occupancy. An example of the reconstructed tracks, per one event,  is given in figure
~\ref{event}.  
    The yield, as reported in the results, is defined as:
    
    \begin{equation}
YIELD(p,\, \theta) =
{N_r^{exp} \over \eta(p, \theta) \cdot A(p, \theta)}
\,,
\end{equation}

     where  $N_r^{exp}$ is the number of measured positrons given by reconstruction,
	    $A(p, \theta)$, the acceptance of the Drift Chamber in the horizontal and
	       vertical planes.
    The acceptance is also derived from the simulations.	       

\begin{figure}
\includegraphics[width=.5\textwidth]{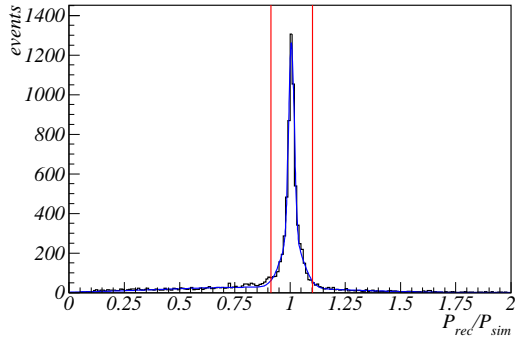}
\caption{Ratio of reconstructed ($P_{rec}$) and emitted ($P_{sim}$) momenta of the positrons
\label{p_ratio}}
\end{figure}

\begin{figure}
\includegraphics[width=.5\textwidth]{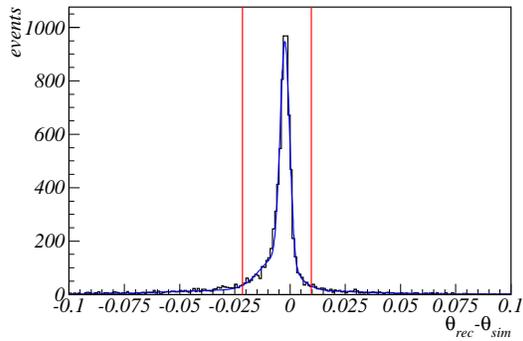}
\caption{Difference of reconstructed ($\theta_{rec}$) and emitted ($\theta_{sim}$) angle of the positrons. Angular
scale is in radians.
\label{a_ratio}}
\end{figure}

Among the reconstructed tracks only a part can be considered as ``good
tracks'' corresponding to a proper evaluation of the positron momentum and
angle. An illustration of the quality of track reconstruction is given in
figures ~\ref{p_ratio} and ~\ref{a_ratio}, which result from simulations,
where the vertical cuts limit the domain of good tracks. Outside of these
cuts, there are ''bad'' reconstructed tracks with poor resolution 
in momentum or
also fake tracks. Assuming that the contamination of bad tracks in the
simulations and in the experiment is the same, this contamination is
automatically corrected in equation (2).

\begin{figure}
\includegraphics[width=.5\textwidth]{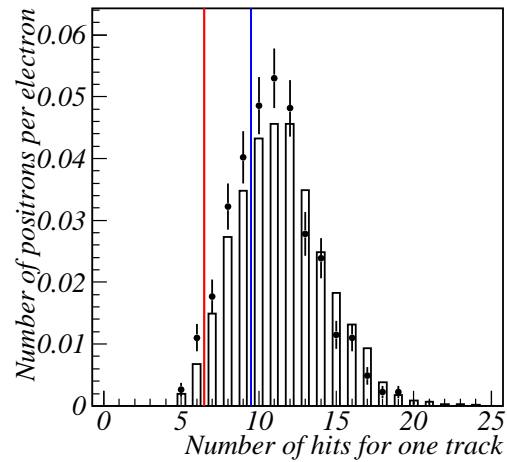}
\caption{Distribution of the number of hits per track, for the 20 mm target, the electron
energy is 10 GeV, the magnetic field is 1 KG. The points are the
experimental data and the histogram is the simulation. Lines show the 2
different cuts $N_{hits}\geq 10$ and $N_{hits}\geq7$ used in the analysis and
systematic errors study.
\label{hits}}
\end{figure}
	       
\begin{figure}
\includegraphics[width=.5\textwidth]{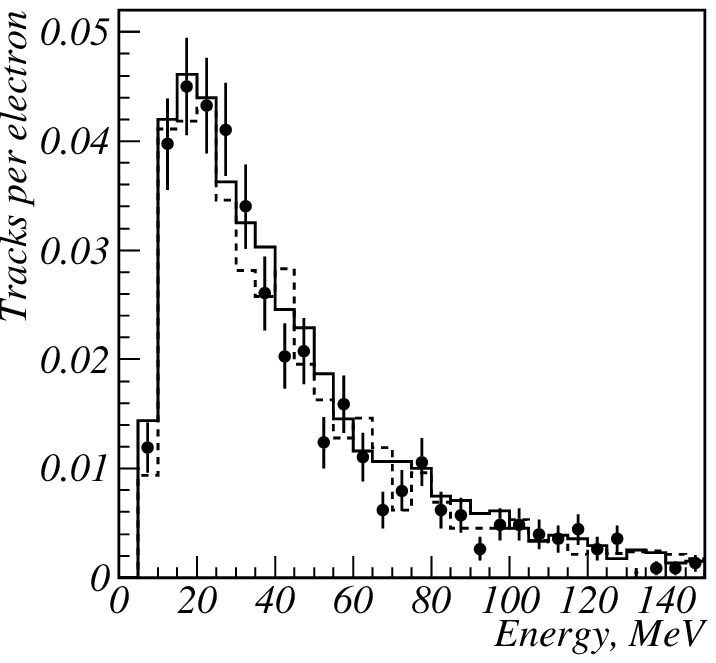}
\caption{The number of measured positrons $N_r^{exp}$  
normalised per 1 incident electron for a  
20 mm thick target. The electron energy is 10 GeV. 
The points with error bars
are the experimental data. 
The 2 histograms are the spectra simulated with GEANT (continuous line)
and SGC (dashed line) generators. No correction for reconstruction efficiency
and acceptance.
\label{x2x10}}
\end{figure}

\subsection{Data selection}

The experimental results were obtained using the following procedure.
\begin{itemize}
\item The incident electrons were selected by the trigger counters.
Additional off-line selection was made with the profile monitor
(Fig.~\ref{delay}). 

\item For further analysis only tracks with a positron-like 
curvature were selected.

\item To reduce the background from the charged particles created outside the target 
(e.g. in the target supporting structure, in the drift chamber walls, \textit{etc.})
the reconstructed track position on the target plane $x_t$ is limited 
within $\pm3\sigma$ around the peak values, where $\sigma$ 
is the RMS of the reconstructed position distribution.

\item To reduce the rate of the fake tracks only tracks with the number of hits
$N_{hits}\ge10$ were selected. In Fig.~\ref{hits} one can see that there is a reasonable 
agreement between the simulation and experimental data for the distribution of
the number of hits 
for one track; 
hence simulation can be used safely to determine the efficiency.

\item The positron registration efficiency was determined 
for each energy and angular bin
using the
simulation. Such efficiency value is needed to evaluate the actual
yield at the target exit. The shower development in the targets was
simulated using an event generator ``Shower Generation in Crystal'' (SGC)
taking into account the specific character of electromagnetic interactions in axially aligned
 crystals. The main features of SGC are given in \cite{BKS4},\cite{BS1}. Note that SGC can be
 used to describe the shower development in amorphous targets as well.

A full description of the detector uses a GEANT code configuration.

The results on positrons presented hereafter are, henceforth, taking into
account the reconstruction efficiency, i.e. the number of reconstructed
positrons for each energy or angle bin was divided by the corresponding
efficiency and detector acceptance. 

The momentum resolution is depending on $p$ value. It is between 2 and 5~\%
for the tracks in the domain of interest ($p < 150$ MeV/c; $\theta < 30$ degrees);  
 the best resolution being for low momenta due to the larger curvature of the
track. 
This momentum resolution has been determined from the comparison between the reconstructed and simulated momenta 
(see Fig.~\ref{p_ratio}).
The value of the momentum resolution has low impact on the shape of
the measured histograms due to the larger energy bin width (5 MeV) with respect 
to this resolution. The momentum resolution is also considered in the
simulations.

\item The detection efficiency, determined by simulations, was obtained for
both values 
of the magnetic field. The data corrected by efficiency  and detector acceptance
for 1 and 4 KG magnetic field values were 
merged and the difference between the data and the simulations 
for the 20 mm amorphous target
serves as an
estimator of systematic errors (see section~\ref{systematic}).

\end{itemize}

\section{Results}

The experimental results on the 20 mm amorphous target and on the 4 and 8 mm
crystals in random orientation have been compared to the simulation with the GEANT 
based generator
and to those obtained with the generator SGC. A good
agreement was met between experiment and simulations in all the cases. That
means that the apparatus and the reconstruction procedure worked quite well.
Moreover, a good agreement was verified between GEANT and the SGC
for the case of the amorphous target case (see Fig.~\ref{x2x10}).

\subsection{Enhancement in photon production}

\begin{figure}
\includegraphics[width=.5\textwidth]{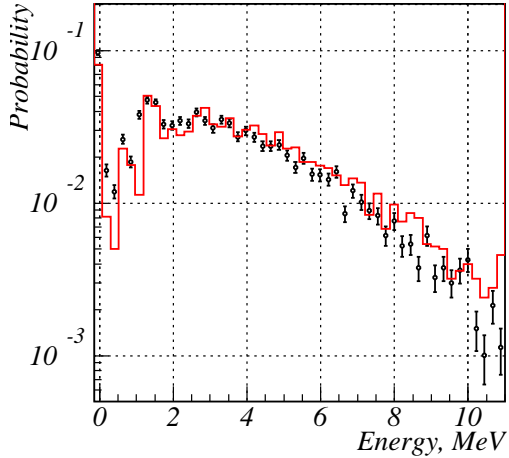}
\caption{The energy loss spectrum in the preshower for the 4 mm crystal target, 
the incident electron energy is 10 GeV, for the converter with 6 cm diameter.
Points represent the experimental data, the histogram represents the simulation.
\label{photon1}}
\end{figure}

\begin{figure}
\includegraphics[width=.5\textwidth]{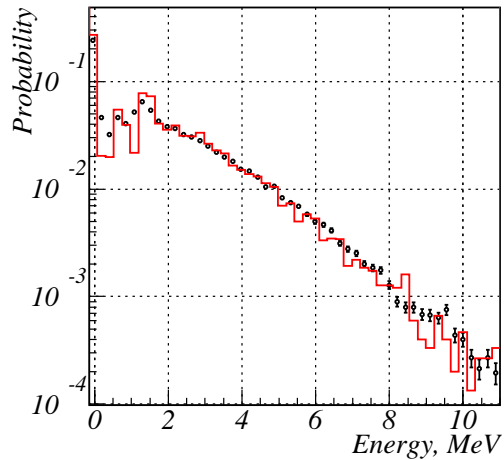}
\caption{The energy loss spectrum in the preshower for the 8 mm crystal target, 
the incident electron energy is 10 GeV, for the converter with 3 cm diameter.
Points represent the experimental data, the histogram represents the simulation.
\label{photon2}}
\end{figure}

The preshower gives the average photon multiplicity in a forward cone as 
a function of the
crystal orientation. On the $<111>$ axis of the tungsten crystal, the
ultrarelativistic electrons radiate more photons than by conventional
bremsstrahlung, as shown on the right rocking curve of figure~\ref{scan}.
Correspondingly, the positron yield is enhanced (see left rocking curve of
Fig.~\ref{scan}).
In Figs.~\ref{photon1}--\ref{photon2} we can see a good agreement between
simulation and experiment concerning the preshower for 4 and 8 mm crystal, the
incident electron energy being 10 GeV. In these figures we represent the
spectrum of the energy lost in the scintillator by the charged particles,
created by the photons in the converter. The first three peaks are
corresponding to i) no interaction, ii)  Compton scattering  and iii) pair
creation. All the results presented in
Figs.~\ref{photon1}--\ref{photon2} are normalized for one event.
These results do not allow to extract the experimental photon spectrum,
but serve as a check of the correctness for the photon part of the SGC generator.

\subsection{Enhancement in positron production}
\begin{figure}
\includegraphics[width=.5\textwidth]{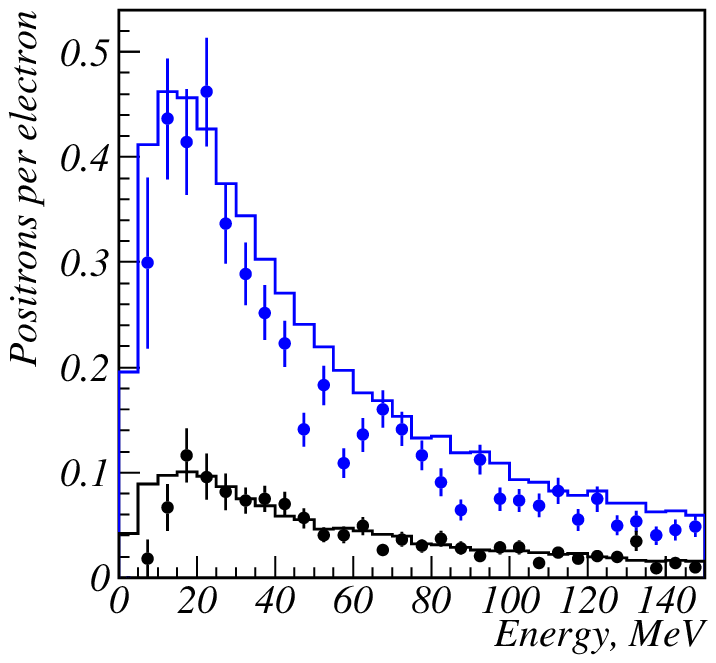}
\hfil
\includegraphics[width=.5\textwidth]{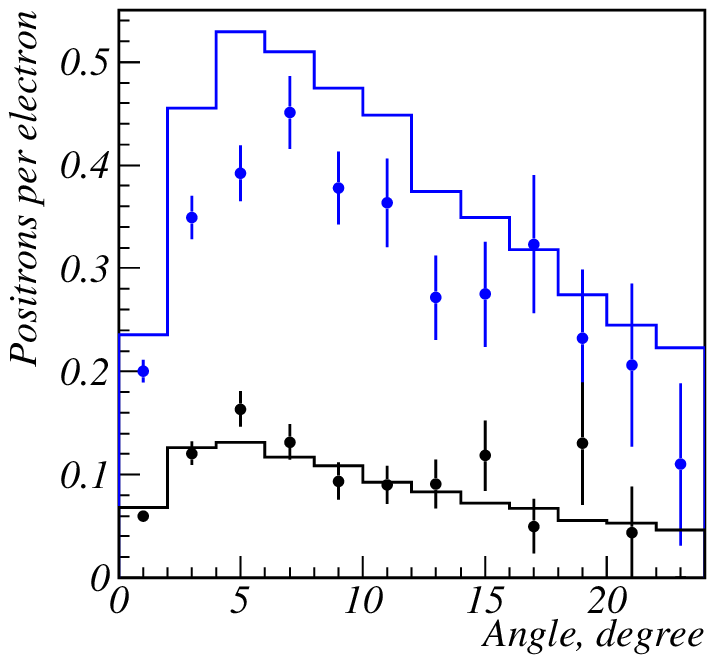}
\caption{The positrons horizontal momentum $p_h$ (left) 
and angular (right) distributions 
for one incident electron and 
4 mm thick target. The electron energy is 10 GeV. 
The points with error bars
are the experimental data. The histograms are the simulated spectra. 
The upper histograms and points on the plots
correspond to the aligned crystal, the bottom histograms and points to the
 random crystal orientation. These distributions are corrected by the reconstruction efficiency
and the detector acceptance.
\label{4mm}}
\end{figure}

\begin{figure}
\includegraphics[width=.5\textwidth]{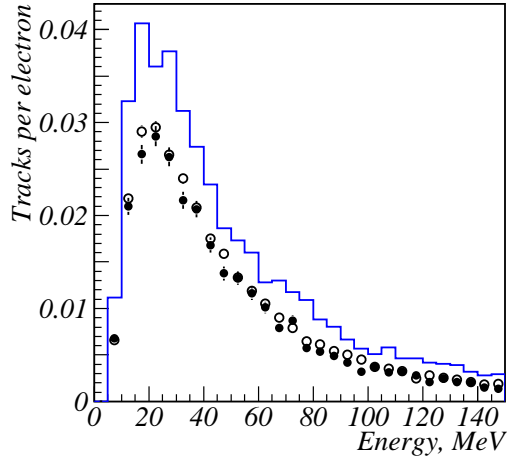}
\caption{The positrons energy spectra for the 1 KG
 magnetic field normalised per 1 incident electron. 
The spectra are not corrected by the reconstruction efficiency and the detector acceptance.
The dark points represent the 8 mm crystal target. The open points,  
the ``4 mm crystal target + 4 mm amorphous target''.
The histogram is the 8mm crystal simulation.
The electron energy is 10 GeV.
\label{x8x10}}
\end{figure}

Typical energy and angular distributions are presented in Fig.~\ref{4mm} for the
4 mm thick crystal and amorphous targets. The energy of the initial electron
is 10 GeV. Simulations and experimental data are presented on the same
picture. 
The energy distribution concerns all
positron angles up to 30 degrees, whereas the angular distribution concerns
all energies more than 5 MeV; that means also high energies (more than 150 MeV) for which 
the discrepancy between simulation and experiment is somewhat larger.
Precisely, these large momenta exhibit low angle values. In the energy
domain of interest, for the positron sources ($5 < p < 50$ MeV) the agreement is
good between simulation and experiment.
Moreover, if we compare the crystal and amorphous targets, an enhancement by
a factor close to 4 is obtained for the oriented crystal. 
Similar considerations may be formulated for the case of the 8 mm crystal
target for which:
\begin{itemize}
\item an enhancement slightly larger than 2 is observed in positron
          production between the crystal and an amorphous target 8 mm thick,
          still for incident electrons of 10 GeV.
          This enhancement is present but slightly lower (than 2) when the
          electron incident energy is of 6 GeV.
\item	  almost identical results about the positron yields are
          obtained for a 8 mm "all crystal" target and a compound target
	  made of a 4 mm crystal followed by a 4 mm amorphous disk. Such results
	  confirm the interest of compound targets for which the first part
          acts as a radiator and the second part
	  as a converter to materialize the photons in $e^+e^-$ pairs. 
\end{itemize}
 Results on the 8 mm target are represented in Fig.~\ref{x8x10}.

\subsection{Presentation of the experimental results in the ($p_L$,$p_T$) space}

\begin{figure}
\includegraphics[width=.5\textwidth]{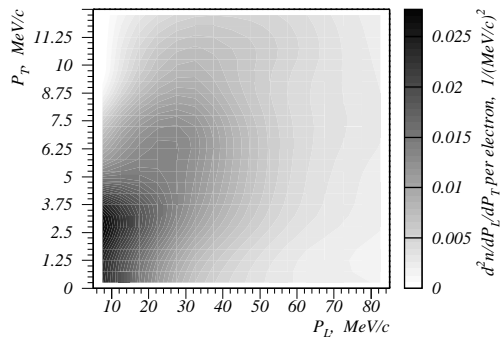}
\caption{The positrons $p_T$ $p_L$ distribution for 8 mm target. 
The incident electron energy is 
10 GeV. The crystal is aligned along $<111>$ axis. 
\label{2d}}
\end{figure}

As the acceptance of the matching systems put after the positron target may 
be defined in the ($p_L$,$p_T$) space with the indication of the maximum accepted
momenta in the longitudinal and transverse directions, we have chosen to
represent the population of the reconstructed positron tracks in such diagrams. Fig.~\ref{2d}
corresponds to the 8 mm crystal target. Results are corrected by the reconstruction efficiency
and the detector acceptance.
It can be seen that the highest densities of positrons are contained in a domain 
defined by $p_L < 20$ MeV/c and $p_T < 5$ MeV/c. Suitable
matching systems (adiabatic devices) which are presently considered for
the future linear colliders \cite{Chehab2} are presenting such kind of acceptance. 

We also present these results, in ($p_L$,$p_T$) space, in the form of tables
(tables \ref{firsttable} to \ref{lasttable}) giving the number of positrons in some relevant ($p_L$,$p_T$)
domains for all the targets.  
In those tables the simulations errors are mainly systematics and found to be near 5\%.
Experimental results are presented with the statistical errors. The numbers are corrected
by the reconstruction efficiency and the detector acceptance.
These
tables where results from experiment and simulations are compared show that:
\begin{itemize}
\item                the expected positron yield may be determined accurately,
                 provided the matching system has been precisely defined.
		 The matching systems are usually characterized by the
		 accepted energy domain and by the maximum transverse
		 momentum; the latter depending on the maximum magnetic
		 fields of the matching lenses.
\item	       the increase of the positron yield with the maximum
                longitudinal momentum for a given
	         maximum transverse momentum is rather slow, showing
		 that there is a predominance of soft positrons
		 with crystal targets.
\item            the accepted yield  may be improved significantly by
	         increasing the maximum transverse momentum. For a given
                 energy domain, $5<p_L<25$ MeV/c the yield is improved by a
		 factor between 1.6 and 2 when increasing the maximum 
		 transverse momentum from 4 to 8 MeV/c. 
\item	        the discrepancies between simulations and experiment are
	         within 10 to 25 \% (for the worst case, i.e. the 8 mm 
		 target). This is not negligible but comparable with
		 usual differences between phase space simulations and 
		 measurements.
\end{itemize}

\subsection{Systematic errors}
\label{systematic}
Various sources of systematic errors on the positron yield 
can be considered. They are discussed below.

Using the simulation,
the influence of the beam angular spread was estimated to be negligible.
The errors induced by the trigger selection were estimated applying a
modified cut on the profiles given by the delay chamber: the error is 3\%.  

Systematic errors on the calibration ``time-radius'' of the drift chamber led to a 3\%
error on the results.

Measurements of the magnetic field contain errors which led, according to
simulations, to an uncertainty of 4\% on the positron momentum measurement.

Comparing the amorphous target simulations for the SGC and GEANT generators, a
 5\% systematic error was assigned to the generator induced uncertainties.

The main source of the systematic errors is in the estimations 
of the efficiency and
fake track rate. This uncertainty has several sources: 
a difference between the real 
and assumed positions of the positron detector relative to the target; 
the charge collection dependence on the track multiplicity in the cell;
the dependence of the gas multiplication factor on the coordinates along 
the wire; border effects in the drift chamber cell.
Comparing the experimental results concerning the 20 mm amorphous target with
the  
simulations using GEANT and SGC generators, the error was found to be 
equal to 8\%.

The total systematic error is  10\%.

Additional reliability checks were performed.
It was checked that the results 
obtained with 1 and 4 KG magnetic field  are the same in the
region of the positron momentum available for both values.

The 20 mm amorphous target gives approximately the same occupancy in the
positron detector as the 8 mm crystal target. The positron spectra for the
20 mm are in good agreement with the simulation, which proves the correct 
efficiency and fake track rate estimation for the case with the biggest 
track occupancy in the experiment.
However it  was found that the 8 mm oriented crystal produces 20\% less
positrons 
compared to the simulation prediction (Fig.~\ref{x8x10}). This could be the indication that the 
simulation
in this case is less precise or that the experimental systematic errors are 
underestimated.

\section{DISCUSSION AND CONCLUSION}

 The main results obtained in this experiment show:
\begin{itemize}
\item      a clear enhancement in photon and positron generations when
        comparing crystal and amorphous targets of the same thickness.
	This enhancement is close to 4 for 4 mm target and larger than 2 for
        8 mm target with a 10 GeV electron beam. The enhancement is
	somewhat lower at 6 GeV.
\item a good equivalence between the results for 8 mm crystal and the
        compound target allows the use of the latter instead of the
        all-crystal, for thermal reasons. 	
\item a large number of soft positrons due to the materialization of
        soft photons. The channeling and coherent bremsstrahlung
        processes provide softer photons than classical bremsstrahlung.
	So an enhanced number of soft positrons due to the materialization of
        these soft photons is available.
	The practical interest is a better matching with known energy 
	acceptances of currently used matching systems.
\item a good agreement between simulations and experiment for photons and
        positrons. This result validates the simulations based on crystal 
 	processes, permitting reliable simulations with different incident
        energies and  crystal types (material, orientation, thickness). That is
        essential to optimize this kind of source.

We may add an additive argument for the use of crystal targets, 
due to investigations on the deposited
energy in these targets \cite{ACCS}. It results from these
calculations that even if the PEDD in crystal and amorphous
targets giving the same yield is practically the same, the total deposited
energy is much less in the crystal case. This set of arguments make this kind 
of positron source attractive. It justifies also the continuous interest in
investigating such sources as it can be verified presently at KEK
\cite{Suwada},\cite{Satoh}.
\end{itemize}

\section{Acknowledgments}

The authors are indebted to Professors J.Lefrancois, F.Richard, A.Skrinsky
for their appreciable support. We are grateful for the technical support in
BINP, LAL and IPNL.

\listoffigures

\begin{table*}
\begin{tabular}[t]{|c||c|c|c|}
\hline
experiment          & $5<p_L<25$ MeV/c   & $5<p_L<30$ MeV/c   & $5<p_L<40$ MeV/c   \\
\hline          
$p_T<4$  MeV/c & $1.16\pm0.04$ & $1.28\pm0.04$ & $1.43\pm0.04$ \\
$p_T<6$  MeV/c & $1.66\pm0.05$ & $1.85\pm0.05$ & $2.13\pm0.05$ \\
$p_T<8$  MeV/c & $2.11\pm0.07$ & $2.46\pm0.08$ & $2.90\pm0.08$ \\
$p_T<10$ MeV/c & $2.31\pm0.08$ & $2.75\pm0.08$ & $3.32\pm0.08$ \\
$p_T<12$ MeV/c & $2.40\pm0.08$ & $2.94\pm0.09$ & $3.67\pm0.10$ \\
\hline          
\hline
simulation          & $5<p_L<25$ MeV/c    & $5<p_L<30$ MeV/c    & $5<p_L<40$ MeV/c    \\
\hline          
$p_T<4$   MeV/c & $1.34$ & $1.49$ & $1.69$ \\
$p_T<6$   MeV/c & $2.06$ & $2.32$ & $2.72$ \\
$p_T<8$   MeV/c & $2.56$ & $2.94$ & $3.51$ \\
$p_T<10$  MeV/c & $2.83$ & $3.30$ & $4.03$ \\
$p_T<12$  MeV/c & $2.93$ & $3.49$ & $4.35$ \\
\hline          
\end{tabular}          
\caption{\label{firsttable}Positron yield: 8mm crystal/10 GeV incident energy.
Domains defined in longitudinal $p_L$ and transverse $p_T$ momenta.}
\end{table*}

\begin{table*}

\begin{tabular}[t]{|c||c|c|c|}
\hline
experiment          & $5<p_L<25$ MeV/c   & $5<p_L<30$ MeV/c    & $5<p_L<40$ MeV/c    \\
\hline          
$p_T<4$   MeV/c & $0.73\pm0.04$ & $0.82\pm0.04$ & $0.92\pm0.04$ \\
$p_T<6$   MeV/c & $1.09\pm0.05$ & $1.24\pm0.05$ & $1.43\pm0.05$ \\
$p_T<8$   MeV/c & $1.34\pm0.07$ & $1.59\pm0.07$ & $1.88\pm0.08$ \\
$p_T<10$  MeV/c & $1.47\pm0.08$ & $1.80\pm0.08$ & $2.19\pm0.08$ \\
$p_T<12$  MeV/c & $1.57\pm0.09$ & $1.99\pm0.09$ & $2.49\pm0.10$ \\
\hline          
\hline
simulation          & $5<p_L<25$ MeV/c    & $5<p_L<30$ MeV/c    & $5<p_L<40$ MeV/c    \\
\hline          
$p_T<4$   MeV/c & $0.90$ & $0.99$ & $1.13$ \\
$p_T<6$   MeV/c & $1.37$ & $1.54$ & $1.79$ \\
$p_T<8$   MeV/c & $1.71$ & $1.96$ & $2.34$ \\
$p_T<10$  MeV/c & $1.90$ & $2.22$ & $2.70$ \\
$p_T<12$  MeV/c & $1.99$ & $2.37$ & $2.94$ \\
\hline          
\end{tabular}          
\caption{Positron yield: 8mm crystal/6 GeV incident energy.
Domains defined in longitudinal $p_L$ and transverse $p_T$ momenta.}
\end{table*}

\begin{table*}

\begin{tabular}[t]{|c||c|c|c|}
\hline
experiment          & $5<p_L<25$ MeV/c    & $5<p_L<30$ MeV/c    & $5<p_L<40$ MeV/c    \\
\hline          
$p_T<4$   MeV/c & $0.71\pm0.06$ & $0.77\pm0.06$ & $0.89\pm0.06$ \\
$p_T<6$   MeV/c & $0.96\pm0.08$ & $1.10\pm0.08$ & $1.31\pm0.08$ \\
$p_T<8$   MeV/c & $1.12\pm0.09$ & $1.38\pm0.10$ & $1.69\pm0.11$ \\
$p_T<10$  MeV/c & $1.14\pm0.09$ & $1.43\pm0.11$ & $1.87\pm0.12$ \\
$p_T<12$  MeV/c & $1.15\pm0.09$ & $1.47\pm0.11$ & $1.94\pm0.13$ \\
\hline          
\hline
simulation          & $5<p_L<25$ MeV/c    & $5<p_L<30$ MeV/c    & $5<p_L<40$ MeV/c    \\
\hline          
$p_T<4$   MeV/c & $0.61$ & $0.69$ & $0.83$ \\
$p_T<6$   MeV/c & $0.87$ & $1.01$ & $1.24$ \\
$p_T<8$   MeV/c & $1.04$ & $1.23$ & $1.55$ \\
$p_T<10$  MeV/c & $1.07$ & $1.30$ & $1.66$ \\
$p_T<12$  MeV/c & $1.07$ & $1.33$ & $1.71$ \\
\hline          
\end{tabular}
\caption{Positron yield: 4 mm crystal/10 GeV incident energy.
Domains defined in longitudinal $p_L$ and transverse $p_T$ momenta.}
\end{table*}

\begin{table*}

\begin{tabular}[t]{|c||c|c|c|}
\hline
experiment          & $5<p_L<25$ MeV/c    & $5<p_L<30$ MeV/c    & $5<p_L<40$ MeV/c    \\
\hline          
$p_T<4$   MeV/c & $0.43\pm0.03$ & $0.47\pm0.03$ & $0.54\pm0.03$ \\
$p_T<6$   MeV/c & $0.66\pm0.05$ & $0.76\pm0.05$ & $0.90\pm0.05$ \\
$p_T<8$   Me
V/c & $0.86\pm0.07$ & $1.01\pm0.07$ & $1.19\pm0.07$ \\
$p_T<10$  MeV/c & $0.93\pm0.07$ & $1.11\pm0.07$ & $1.33\pm0.08$ \\
$p_T<12$  MeV/c & $0.96\pm0.07$ & $1.17\pm0.08$ & $1.45\pm0.08$ \\
\hline          
\hline
simulation          & $5<p_L<25$ MeV/c    & $5<p_L<30$ MeV/c    & $5<p_L<40$ MeV/c    \\
\hline          
$p_T<4$   MeV/c & $0.47$ & $0.53$ & $0.63$ \\
$p_T<6$   MeV/c & $0.71$ & $0.81$ & $0.98$ \\
$p_T<8$   MeV/c & $0.87$ & $1.01$ & $1.24$ \\
$p_T<10$  MeV/c & $0.92$ & $1.09$ & $1.37$ \\
$p_T<12$  MeV/c & $0.93$ & $1.13$ & $1.43$ \\
\hline          
\end{tabular}          
\caption{Positron yield: 4 mm crystal/6 GeV incident energy.
Domains defined in longitudinal $p_L$ and transverse $p_T$ momenta.}
\end{table*}

\begin{table*}
\begin{tabular}[t]{|c||c|c|c|}
\hline
experiment          & $5<p_L<25$ MeV/c    & $5<p_L<30$ MeV/c    & $5<p_L<40$ MeV/c    \\
\hline
$p_T<4$  MeV/c  & $1.62\pm0.15$ & $1.74\pm0.15$ & $1.89\pm0.16$ \\
$p_T<6$  MeV/c  & $2.46\pm0.22$ & $2.63\pm0.22$ & $2.93\pm0.22$ \\
$p_T<8$  MeV/c  & $2.91\pm0.24$ & $3.33\pm0.26$ & $3.77\pm0.26$ \\
$p_T<10$ MeV/c  & $3.17\pm0.25$ & $3.76\pm0.28$ & $4.37\pm0.29$ \\
$p_T<12$ MeV/c  & $3.41\pm0.32$ & $4.08\pm0.34$ & $4.82\pm0.36$ \\
\hline
\hline
simulation          & $5<p_L<25$  MeV/c   & $5<p_L<30$  MeV/c   & $5<p_L<40$  MeV/c   \\
\hline
$p_T<4$  MeV/c  & $1.39$ & $1.51$ & $1.68$ \\
$p_T<6$  MeV/c  & $1.99$ & $2.22$ & $2.53$ \\
$p_T<8$  MeV/c  & $2.49$ & $2.83$ & $3.29$ \\
$p_T<10$ MeV/c  & $2.71$ & $3.14$ & $3.74$ \\
$p_T<12$ MeV/c  & $2.80$ & $3.29$ & $3.94$ \\
\hline
\end{tabular}
\caption{\label{lasttable}Positron yield: 20 mm amorphous/10 GeV incident energy.
Domains defined in longitudinal $p_L$ and transverse $p_T$ momenta.}
\end{table*}

\end{document}